\documentclass[amsmath,amssymb,groupedaddress,lengthcheck,aps,prb]{revtex4}

\usepackage{graphics}
\usepackage{color}
\usepackage{graphicx}
\usepackage{bm}
\usepackage[latin1]{inputenc}
\usepackage{textcomp}
\renewcommand{\vec}[1]{\mbox{\boldmath$#1$}}
\def\be{\begin{equation}}
\def\ee{\end{equation}}

\begin{document}

\title{A note contrasting two microscopic theories of the fractional quantum Hall effect}

\author{Jainendra K. Jain}
\affiliation{Physics Department, 104 Davey Laboratory, Pennsylvania State University, University Park, PA 16802}

\begin{abstract}
Two microscopic theories have been proposed for the explanation of the fractional quantum Hall effect, namely the Haldane-Halperin hierarchy theory and the composite fermion theory. Contradictory statements have been made regarding the relation between them, ranging from their being distinct to their being completely equivalent. This article attempts to provide a clarification of the issue. It is shown that the two theories postulate distinct  microscopic mechanisms for the origin of the fractional quantum Hall effect, and make substantially different predictions that have allowed experiments to distinguish between them.
\end{abstract}

\maketitle

\section{Introduction}

The first fractional quantum Hall effect\cite{Tsui82} (FQHE) observed in 1982 was labeled by the fraction 1/3. It was explained by Laughlin\cite{Laughlin83} in 1983 by construction of wave functions for the ground states and their quasiparticles at Landau level fillings of $\nu=1/m$, where $m$ is an odd integer. Immediately thereafter some non-$1/m$ fractions were observed. For explaining their origin, Haldane and Halperin put forward in 1983 - 1984 a hypothesis known as the Haldane-Halperin (HH) hierarchy theory\cite{Haldane83,Halperin84}. As the sample quality was improved, a huge number of additional FQHE states were revealed, which inspired another hypothesis for the origin of the FQHE, called the composite fermion (CF) theory\cite{Jain89,Lopez91,Halperin93,Stormer95,PT93,Jain07,Heinonen98,Jain98,Lopez98,Simon98,DasSarma96,Halperin96,Stormer96,Jain00,Halperin03,Smet98a,Murthy03}. There appears to be a misconception in the literature regarding the relation between these two theories; specifically, many articles characterize them as being equivalent\cite{Sitko97,Hansson09,Surosa1,Bonderson,Papic13,Kvorning13}, and some even assert\cite{Hansson09,Surosa1,Bonderson} that the CF theory is a special case of, and can be microscopically derived from, the HH hierarchy theory.
This note is intended as an attempt to clarify the issue, and to show that the conflation of the two theories is incorrect. They attribute distinct, and competing, physical mechanisms to the origin of the FQHE and are distinguishable through their substantially different experimental predictions.  

Both theories are hierarchical in the sense that they both arrange FQHE states in order of importance. They both postulate candidate states at all odd denominator fractions (and also at even denominator fractions in the CF theory). They yield identical fractional values for the local charge\cite{Laughlin83} and the braid statistics\cite{Halperin84} of the quasiparticle excitations. Consequently, they belong to equivalent effective K-matrix Chern-Simons actions\cite{CS}, which encode information on these fractional quantum numbers. The effective K-matrix action has been used as a starting point by topological field theorists who are only interested in certain universal asymptotic low energy properties of a FQHE state rather than its microscopic origin.

It is important to understand, however, that an agreement on these aspects does not imply an equivalence of the two theories. The K-matrix formulation is not a microscopic theory but an ``effective" description that is designed to deal with certain universal aspects of the FQHE; it can be experimentally tested through its sharp predictions\cite{universal,Chang03}, but it  does not necessarily distinguish between different candidate microscopic theories of the phenomenon. To give an example from another contemporary problem: The effective Landau Ginzburg action with d-wave spin-singlet order parameter can serve as a starting point for describing certain general aspects of the high temperature cuprate superconductors. However, all candidate theories that produce d-wave spin-singlet pairing (based inevitably on approximate treatments) can surely not be deemed equivalent explanations of the microscopic origin of high T$_{\rm c}$ superconductivity. That is the reason why the high T$_{\rm c}$ superconductivity community has been engaged in a vigorous and painstaking process of deducing and testing all of the qualitative and quantitative consequences of various candidate theories.

When a broader set of FQHE facts is considered, the CF and the HH hierarchy theories are seen to differ in important ways that are highlighted in this article. The central feature of the CF theory is the introduction of ``new particles"\cite{Stormer95} called composite fermions, which experience an effective magnetic field. As discussed in Section~\ref{secIII}, composite fermions have been observed experimentally, as have many of their states, phenomena, and other properties. The prominent FQHE at fractions of the form $n/(2pn\pm 1)$, where $p$ and $n$ are integers, has been explained as the integer quantum Hall effect (IQHE) of composite fermions\cite{Jain89}, which gives an intuitive explanation of the relative sizes of the energy gaps. The compressible state at $\nu=1/2$ has been successfully described as a Fermi sea of composite fermions\cite{Halperin93}, which has given rise to much phenomenology. Composite fermions have also suggested a natural way to understand the even denominator 5/2 FQHE state\cite{Willett87} as an instability of the CF Fermi surface to the formation of a p-wave ``superconductor" of composite fermions\cite{Moore91,Read00}. Composite fermions provide an explanation of the role of spin in the FQHE, as well as of various types of charged and neutral excitations.

Further, the CF physics has led naturally to a set of microscopic trial wave functions\cite{Jain89} for the ground states and excitations at all filling factors of the form $\nu=n/(2pn\pm 1)$, by analogy to the known wave functions for the corresponding IQHE states. These have been found to be accurate representations of the actual Coulomb eigenstates of interacting electrons in the lowest Landau level, and have allowed detailed quantitative comparisons with experiments (Sec.~\ref{secIII}). The Chern-Simons field theoretical formulation of composite fermions\cite{Lopez91,Halperin93}, with a proper analysis of the interactions mediated by the Chern-Simons gauge field, enables one to describe\cite{Halperin93} dynamics and transport phenomena at $\nu=1/2$. 

The HH hierarchy theory does not contain composite fermions and the physics arising from them. It instead proposes to generate new daughter FQHE states from a Laughlin-like FQHE of the Laughlin-like quasiparticle excitations of a given parent state, and obtains candidate FQHE states at all odd denominator fractions by an iteration of this process starting from the $1/m$ state. The differences between the experimental manifestations of the two approaches are discussed below.

The article is organized as follows. It begins with a brief review of the basic postulates of the HH hierarchy theory and the CF theory (Sec.~\ref{secI}), lists their principal similarities and differences (Sec.~\ref{simdif}), reviews experimental facts (Sec.~\ref{secII}), addresses the universal aspects (Sec.~\ref{secIII}), and concludes (Sec.~\ref{conclusion}). Some parts of the discussion below should be well known to the experts, but are included here for completeness -- a reader who either has not kept up with various developments or has entered into the field more recently may find them useful. The account below is restricted, with the exception of some comments in the last paragraph in Sec.~\ref{conclusion}, to the physics of the lowest Landau level, because that is well understood and suffices for the issue at hand.

\section{Haldane-Halperin hierarchy and composite-fermion theories} 
\label{secI}

A theory of the FQHE needs to address the following interrelated questions: 

\begin{enumerate}

\item {\em What are the ``constituent particles", or building blocks, of the FQHE?}
\item {\em What  is the physical mechanism of incompressibility?} 
\item {\em How do we describe the physics quantitatively?}
\item {\em What are the experimentally testable predictions?}
\end{enumerate}

A remark on the term ``constituent particles" is in order\cite{Halperin}: The most fundamental particles of the FQHE are of course electrons. However, the FQHE is likely to be understood in terms of certain emergent particles, which can be called its constituent particles. The constituent particles of a phenomenon are typically certain weakly interacting entities in terms of which the phenomenon is conveniently described. Consider for example $^4$He superfluidity. One can, in principle, begin with electrons and quarks, but that would not be a useful starting point. An explanation of superfluidity would be impossible without the recognition that quarks and electrons form bound states to produce the emergent $^4$He atoms, which are the constituent particles for the phenomenon of superfluidity. Composite fermions, although more complicated than $^4$He atoms, are the constituent particles of the FQHE in a similar sense.

We briefly review the HH hierarchy and the CF theories in this section.  Both theories consist of two parts: the qualitative physical picture and the quantitative microscopic implementation.

\subsection{\bf Haldane-Halperin hierarchy theory}

\begin{figure}
\vspace{-1.6cm}
\includegraphics[width=3.5in]{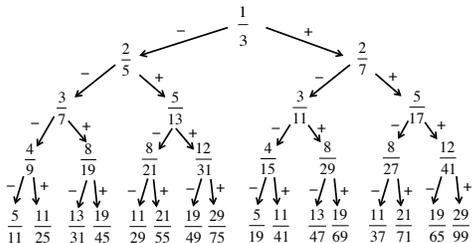}
\vspace{-2.0cm}
\caption{The HH hierarchy tree originating from 1/3. The ``$\pm$" signs correspond to the choices $q=\pm 2$.
\label{HHfrac}}
\end{figure}

\underline{Physical picture}: 
The HH hierarchy theory \cite{Haldane83,Halperin84} begins with the Laughlin $\nu=1/m$ state, $m$ odd, as given. (The electron filling factor $\nu$ is defined as $\nu=\rho hc/eB$, $\rho$ being the electron density.) It further assumes that a change in the filling factor results in the creation of the  ``Laughlin quasiparticles" (LQPs) \cite{Laughlin83}. In what follows, we will denote the electron coordinates by $z=x+iy$ or $\vec{r}=(x,y)$ and the LQP coordinates by $\eta=\eta_x+i\eta_y$ or $\vec{\eta}=(\eta_x,\eta_y)$.  In Haldane's version\cite{Haldane83}, the LQPs are treated nominally as bosons at an effective flux of $\pm N\phi_0$, because each boson sees either a vortex or an anti-vortex at the positions of the $N$ electrons ($\phi_0=hc/e$ is called the flux quantum).  No new incompressible states would be produced if the LQPs were noninteracting, no matter how many LQPs are created. If, however, the interaction between the them is dominated by the short-range repulsive part of the pair interaction and is also weak compared to the energy gap of the parent state\cite{Haldane83}, they can form Laughlin-like {\em fractional} QHE states of their own to produce new incompressible ``daughter" states when $N_{\rm LQP}/N=1/q$, $q$ even integer, which will occur at $\nu=1/(m\pm q^{-1})$. If a daughter exists, and if its { own Laughlin-like quasiparticles} have an appropriate interaction, it can produce granddaughters. Iterating the process gives the fractions 
\be
\nu= \frac{1}{\displaystyle m 
     \pm \frac{1}{\displaystyle q_2
     \pm \frac{1}{\displaystyle q_3 
     \pm \frac{1}{\displaystyle q_4
     \pm \frac{1}{\displaystyle \cdots \pm {1\over q_n} }}}}}
     \label{contfrac}
\ee
at the $n^{\rm th}$ generation. With $q_j=2$, which is expected to produce the strongest daughters at each generation, the HH hierarchy family tree contains $2^{n-1}$ fractions at the $n^{\rm th}$ generation. All odd denominator fractions can be generated in this manner.
In Halperin's version\cite{Halperin84}, the quasiparticles are treated as particles with fractional charge\cite{Laughlin83} and fractional braid statistics\cite{Halperin84,Leinaas77}, but the same daughter fractions are obtained as above.  Fig.~\ref{HHfrac} shows the HH hierarchy tree stemming from 1/3.

\underline{Microscopic implementation}: A microscopic implementation of the HH hierarchy requires constructing trial wave function for the daughter states. For the first daughters at filling factor $\nu>1/m$, these are written as 
\begin{equation}
\int \left[   \prod_{\mu=1}^{N/q} d^2 \vec{\eta}_{\mu}\right]
\bar{\Phi}_{\rm 1/q}( \vec{\eta}_\mu )
\Psi^{\rm LQP}_{1/m}( \vec{r}_j ;  \vec{\eta}_{\mu} )
\label{H}
\end{equation}
where $\Phi_{1/q}=\prod_{\mu<\nu}(\bar{\eta}_{\mu}-\bar{\eta}_{\nu})^q$, and $\Psi^{\rm LQP}_{1/m}( \vec{r}_j ;  \vec{\eta}_{\mu} )$ is a wave function for LQPs at the positions $\{ \vec{\eta}_{\mu} \}$. Following Laughlin's ansatz for the quasiparticles of the $\nu=1/m$ states \cite{Laughlin83}, the latter can be taken as
\begin{equation}
\Psi^{\rm LQP}_{1/m}( \vec{r}_j ;  \vec{\eta}_{\mu} )=\prod_{j=1}^N \prod_{\mu=1}^{N_{\rm LQP}}\left(2{\partial \over \partial z_j}-\bar{\eta}_{\mu}\right)\Phi_{1/m}
\label{LQP}
\end{equation} 
Wave functions for the daughter states at $\nu<1/m$ are constructed using Laughlin's quasihole wave function given by $\prod_{j=1}^N \prod_{\mu=1}^{N_{\rm LQH}}(z_j-\eta_{\mu})\Phi_{1/m}$ for $N_{\rm LQH}=N/q$ located at $\{ \vec{\eta}_{\mu} \}$. The wave functions for the subsequent generations are more complicated and not reproduced here. It should be stressed that the choice of the HH hierarchy wave functions is not unique, and other trial wave functions inspired by this physics have also been studied \cite{Morf86,Morf87,Greiter}.

\vspace{4mm}

\subsection{\bf Composite fermion theory} 

\vspace{4mm}

\underline{Physical picture}: 
The starting point of the CF theory \cite{Jain89,Lopez91,Halperin93,Stormer95,PT93,Jain07,Heinonen98,Jain98,Lopez98,Simon98,DasSarma96,Halperin96,Stormer96,Jain00,Halperin03,Smet98a,Murthy03} is the postulate that electrons capture an even number ($2p$) of quantized vortices each; these bound states are called ``composite fermions." It further postulates that the composite fermions can be treated, in a first approximation, as weakly interacting particles. The Berry phases produced by the vortices bound to composite fermions effectively cancel a part of the external magnetic field, and, as a result, composite fermions experience an effective magnetic field 
\begin{equation}
B^*=B-2p\rho\phi_0
\label{B*}
\end{equation}
where $\rho$ is the electron or CF density. The effective magnetic field is a direct consequence of the bound vortices, and is thus the defining property of composite fermions. Composite fermions form CF Landau levels (also called ``$\Lambda$ levels"; these reside within the lowest electron Landau level) in the effective magnetic field, 
and have a filling factor $\nu^*=\rho hc/e|B^*|$ given by
\begin{equation}
\nu={\nu^*\over 2p\nu^*\pm 1}
\label{cfnu}
\end{equation}
where the ``$-$" sign correspond to the situations when $B^*$ and $B$ point in opposite directions.

\begin{figure}\vspace{-1.0cm}
\includegraphics[width=3.0in]{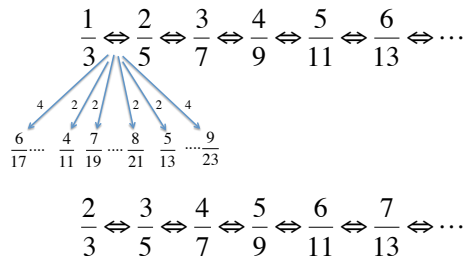}
\vspace{-1.5cm}
\caption{The CF hierarchy of FQHE states arising from composite fermions carrying two vortices. The first level contains the CF IQHE sequences $n/(2n+1)$ and $n/(2n-1)$. The second level consists of new fractions arising from the FQHE of composite fermions; to avoid clutter, only some possible second-level states in the range $2/5>\nu>1/3$ (i.e. $2>\nu^*>1$) are shown. The displayed fractions 6/17, 4/11, 7/19, 8/21, 5/13 and 9/23 at the second level correspond to CF FQHE at $\nu^*=1+/5$, $1+1/3$, $1+2/5$, $1+3/5$, $1+2/3$ and $2-1/5$. The integer near each arrow shows the number of additional vortices the composite fermions in the second CF Landau level must capture to produce the indicated fraction.
\label{cfsequences}}
\end{figure}

Many predictions immediately follow from composite fermions. The IQHE of composite fermions at $\nu^*=n$ manifests as FQHE at the principal sequences 
\begin{equation}
{\nu={n\over 2pn\pm 1}}.
\label{cffrac}
\end{equation}
The states for $2p=2$ are shown in Fig.~\ref{cfsequences}\cite{comm-1m}.
The CF cyclotron energy is given by 
$\hbar\omega^*_{\rm c}={\hbar e|B^*|\over m^* c}\sim {1\over 2pn\pm 1} {e^2\over \epsilon l}$, where we have used $|B^*|={B\over 2pn\pm 1}$, $l=\sqrt{\hbar c/eB}$ is the magnetic length, $\epsilon$ is the dielectric constant of the host material, and the CF effective mass\cite{Halperin93} is taken to be $m^*\propto \sqrt{B}$ to ensure that the CF cyclotron energy has units of the Coulomb interaction (which is the only energy scale in a theory confined to the lowest Landau level).  The fully spin polarized states at $\nu=1-n/(2pn\pm 1)$ are understood by particle hole symmetry; equivalently, one can view these states as $\nu^*=n$ IQHE states of composite fermions created from binding vortices to holes in the lowest Landau level.

The analogy to the IQHE also gives a physical picture for the excitations. A charged excitation is an isolated composite fermions in an otherwise empty CF Landau level or a missing composite fermion from a full CF Landau level, and a neutral excitation is a particle hole pair, or an exciton, of composite fermions. 

Halperin, Lee and Read made a pivotal prediction\cite{Halperin93} that composite fermions form a Fermi sea at $\nu=1/2p$, the $n\rightarrow\infty$ limit of the sequence $n/(2pn\pm 1)$, where the effective magnetic field experienced by them vanishes ($B^*=0$).

The CF Landau level physics leads to predictions for non-fully spin polarized FQHE. For spinful composite fermions, we write $\nu^*=\nu^*_\uparrow+\nu^*_\downarrow$, where $\nu^*_\uparrow$ and $\nu^*_\downarrow$ are the filling factors of up and down spin composite fermions. The possible spin polarizations of the various FQHE states are then predicted by analogy to the IQHE of spinful electrons. For example, the 4/7 state maps into $\nu^*=4$, where we expect, from a model that neglects interaction between composite fermions, a spin singlet state at very low Zeeman energies (with $\nu^*=2+2$), a partially spin polarized state at intermediate Zeeman energies ($\nu^*=3+1$), and a fully spin polarized state at large Zeeman energies ($\nu^*=4+0$). { The model of noninteracting composite fermions  further predicts, with the CF mass as a single adjustable parameter (the appropriate CF mass in this context is different from that defined above due to renormalization by exchange interaction)}, the critical Zeeman energies where transitions from one spin polarization to another take place through crossings of CF Landau levels of different spins, as well as the spin polarization of the CF Fermi sea as a function of the Zeeman energy or the temperature. For non-fully spin polarized states, particle hole symmetry relates $\nu=n/(2pn\pm 1)$ to $\nu=2-n/(2pn\pm 1)$.

The fractions given in Eq.~\ref{cffrac} occur for noninteracting composite fermions, and belong to the first level of the CF hierarchy. One can consider the possibility that the residual interaction between composite fermions can give rise to new states arising from a {\em fractional} QHE of composite fermions. These states would constitute the second level of the CF hierarchy. Some fractions at the second level are shown in Fig.~\ref{cfsequences}. More fractions can be obtained at third and further levels, but are unlikely to be relevant for the Coulomb interaction.

\underline{Microscopic implementation}: To perform quantitative calculations of the gaps, effective masses, response functions, exciton dispersion, spin phase diagram, etc., we need microscopic formulations of the CF physics. Several complementary formulations have been developed. These include the fermion Chern-Simons field theory of Lopez-Fradkin\cite{Lopez91} and Halperin-Le-Read\cite{Halperin93}, and the Hamiltonian theory of Murthy and Shankar\cite{Murthy03}. Another formulation is to construct microscopic { trial wave functions} based on the CF mapping between the FQHE and the IQHE\cite{Jain89}. The wave functions for the ground states at $\nu=n/(2pn+1)$ are constructed as
\begin{equation}
\Psi^{\rm gs}_{\nu={n\over 2pn+1}}=\Psi^{\rm CF, gs}_{\nu^*=n}={\cal P}_{\rm LLL}\Phi^{\rm gs}_{\nu^*=n} \prod_{j<k}(z_j-z_k)^{2p}
\label{Psi_n}
\end{equation}
Here, $\Phi^{\rm gs}_{\nu^*=n}$ is the known electron wave function of $n$ filled Landau levels, the Jastrow factor $\prod_{j<k}(z_j-z_k)^{2p}$ attaches $2p$ vortices to electrons to transform them into composite fermions, ${\cal P}_{\rm LLL}$ projects the wave function into the lowest Landau level, and the right hand side is interpreted as the wave function of $n$ filled composite fermions Landau levels, denoted as $\Psi^{\rm CF, gs}_{\nu^*=n}$.  These wave functions contain no adjustable parameter. For $\nu^*=1$, Eq.~\ref{Psi_n} recovers the Laughlin wave function (with $m=2p+1$), which is interpreted as one filled CF Landau level of composite fermions carrying $2p$ quantized vortices. The wave function for $\nu=n/(2pn-1)$ is produced by replacing $\Phi^{\rm gs}_{\nu^*=n}$ by its complex conjugate.  

The CF theory gives wave functions for charged and neutral excitations at all fractions $n/(2pn\pm 1)$, and also for partially spin polarized or spin singlet FQHE states and their charged and neutral excitations, by analogy to the known wave functions of the corresponding states of noninteracting electrons at $\nu^*$; these also do not involve any adjustable parameters. At $\nu=\nu^*/(2p\nu^*\pm 1)$ with $\nu^*\neq$ integer, the physics is determined by the weak residual interaction between composite fermions, and we can consider various states of composite fermions (e.g. their FQHE, crystals, stripes, Fermi sea, paired state) and construct wave functions for them by analogy to the wave function of interacting electrons at $\nu^*\neq$ integer. Alternatively, we can diagonalize the Coulomb interaction in the low energy CF basis\cite{Mandal02}  
\begin{equation}
\{\Psi_{\nu={\nu^*\over 2p\nu^*+1}}^{a}\}=\{\Psi_{\nu^*}^{\rm CF, a}\}=\{{\cal P}_{\rm LLL}\Phi_{\nu^*}^{a} \prod_{j<k}(z_j-z_k)^{2p}\} 
\end{equation}
where \{$\Phi_{\nu^*}^{a}$\} is the set of lowest kinetic energy wave functions (labeled by ``a") of non-interacting electrons at $n\leq \nu^* \leq n+1$ in which $n$ Landau levels are fully occupied and the $(n+1)$th Landau level is partially occupied.

%


\begin{center}
\begin{figure*}
\includegraphics[width=5.0in]{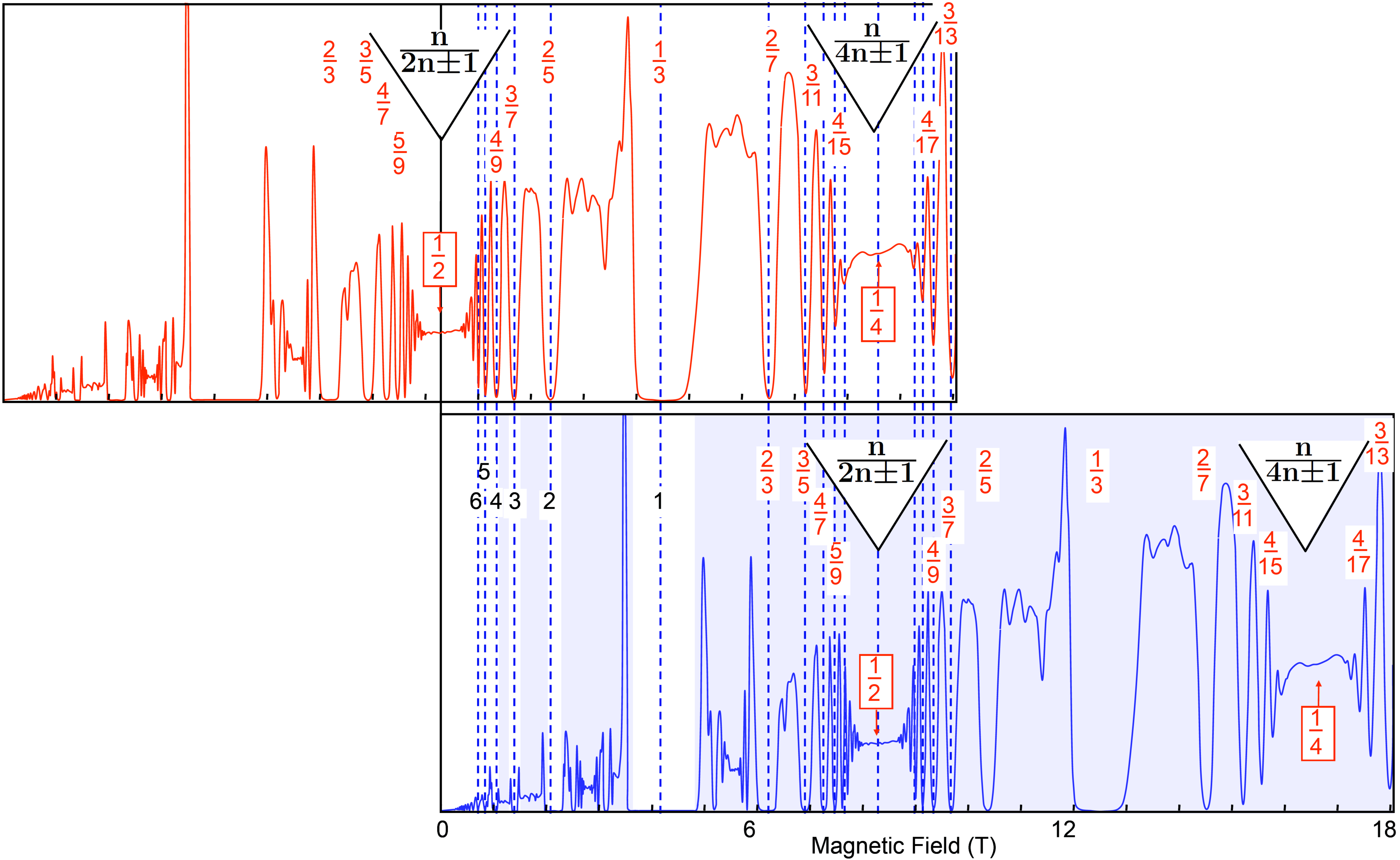}
\caption{Comparing the FQHE data plotted as a function of $B^*=B-2\rho\phi_0$ (upper panel) with the IQHE of electrons (lower panel). 
A close correspondence between the fractions $\nu=n/(2n\pm 1)$ and the integers $\nu^*=n$ is evident. The fractions $\nu=n/(4n\pm 1)$ will map into integers when the upper panel is plotted against $B^*=B-4\rho\phi_0$, the magnetic field seen by composite fermions carrying four vortices. 
Source: H. L. Stormer~\cite{Stormer-IQHE-FQHE}. 
\label{fqhe}}
\end{figure*}
\end{center}

\section{\bf Predictions for experiments: Similarities and differences}
\label{simdif}


Before proceeding to experiments, let us note some similarities and differences between the experimentally measurable consequences of the two theories.

\subsubsection{Composite fermions}

The key signature of the CF theory is that the existence of composite fermions, which experience an effective magnetic field $B^*$ given by Eq.~\ref{B*}.  Composite fermions can be observed through their dynamics, their excitations, and their various many body states. 

The HH hierarchy theory does not contain composite fermions. In particular,  the LQPs and composite fermions are distinct objects.
That should be obvious from the simple fact that non-interacting composite fermions show IQHE and form a Fermi sea, whereas non-interacting LQPs have neither IQHE nor a Fermi sea. That is the reason why in the HH hierarchy theory it is necessary to appeal to a {\em fractional} QHE of the LQPs to envision new incompressible states. Appendix~\ref{ap:LQP} elaborates how the states generated by creating LQPs are distinct from those obtained by adding composite fermions to a CF Landau level.

\subsubsection{Stability of fractions}
\label{stability}

The CF and HH hierarchy theories make different predictions for which fractions are expected to occur prominently in experiments. As noted above, a HH hierarchy daughter results from a Laughlin-like FQHE of the Laughlin-like quasiparticles of the parent. The stabilization of their Laughlin state requires \cite{Haldane83,Quinn00} that the interaction between the quasiparticles be (i) weak compared to the excitation gap of the parent, and (ii) dominated by a short-range repulsion in the pair interaction.  The latter is a nontrivial requirement, because the LQPs are not point particles but have large sizes and a complicated internal structure \cite{comm11}.  (The density profiles for the quasihole and quasiparticle of the 1/3 state are shown, e.g., in Figs.~8.4-8.6 of Ref.~\onlinecite{Jain07}; these have diameters of $\sim$10 and $\sim$12 magnetic lengths, respectively.) If a HH hierarchy daughter is stabilized, it would be substantially weaker than the parent due to the requirement (i) above, i.e. would have a much smaller excitation gap than the parent. The LQPs of the daughter, therefore, would be even larger and more complicated, making the birth of granddaughters even more unlikely.  As a result, from the HH hierarchy perspective, one would expect only a few fractions of the early generations to appear in experiments, with their excitation gaps rapidly decaying from one generation to next. 

The CF theory, on the other hand, predicts prominent FQHE at the sequences $n/(2pn\pm 1)$, with their gaps expected to diminish only as $\hbar\omega^*_{\rm c}\sim {1\over 2pn\pm 1} {e^2\over \epsilon l}$. These fractions lie at the first level of the CF hierarchy. It is possible that, as a result of the weak residual interaction between them, composite fermions might also exhibit FQHE to produce additional states at $\nu\neq n/(2pn\pm 1)$. For the same reason as explained for the FQHE of the LQPs, we expect that the FQHE states of composite fermions, should they occur at all, will be significantly weaker than the IQHE states of composite fermions. If the {\em electron} FQHE is any guide, we should not expect FQHE of composite fermions in high CF Landau levels. Some possible CF-FQHE states in the second CF Landau level (i.e. for $2/5>\nu>1/3$) are shown in Fig.~\ref{cfsequences}.

As a concrete example, take the fraction 6/13. It appears as one of the $2^5$ fractions at the 6th generation of the HH hierarchy. The HH hierarchy states at the 6th generation are not expected to occur, as they would entail stabilization of 5 successive FQHE states on top of one another, formed from 5 different sets of LQPs that are increasingly more complex with each generation. In contrast, 6/13 occurs at the first generation of the CF hierarchy.

\subsubsection{CF Fermi sea}
\label{CFFS}

No CF Fermi sea emerges in the HH hierarchy theory as we go toward $\nu=1/2p$. This is related to the different physical origins assigned to FQHE in the two theories.

\subsubsection{Spin polarization}

The CF Landau level physics leads to predictions for non-fully spin polarized FQHE, resulting from a competition between the CF cyclotron energy and the Zeeman energy.

\subsubsection{Local charge and braid statistics of excitations}
\label{topological}

Laughlin predicted\cite{Laughlin83} that the local charge $e^*$ of the excitations (i.e., charge excess or deficiency relative to the uniform FQHE state) has a $\nu$ dependent fractional value, and Halperin predicted\cite{Halperin84} that they obey fractional braid statistics characterized by a parameter $\theta^*$. The HH hierarchy and CF theories produce the same values for $e^*$ and $\theta^*$. 

\subsubsection{Other experimental observables}

The qualitative differences noted above arise from the ``physical pictures" of the two theories. Further differences will appear through detailed quantitative calculations, based on the ``microscopic implementations" of these theories, for various experimentally measurable quantities, such as ground state energies, gaps, neutral excitations, spin wave modes, spin polarization phase diagram, various response functions, {\em etc}.

\section{Experimental facts}
\label{secII}

{\em Both the HH hierarchy and the CF theories are postulates,} and it is ultimately up to the experiments to decide which of the two, if either, is chosen by nature. Fortunately, a great variety of FQHE facts have been established during the past three decades. There is evidence for $\sim$70 fractions in the lowest Landau level\cite{Stormer-IQHE-FQHE} (with $0<\nu<2$), and states with several different spin polarizations have been observed at many of these fractions. The gaps, spin polarizations, collective mode dispersions, etc. have been measured for many of these FQHE states. Here is a partial list of experimental facts that have been successfully predicted or explained, in a unified fashion, by the CF Theory (for further details and numerous other phenomena, see the reviews in Refs.~\onlinecite{Jain07,Heinonen98,Jain98,Lopez98,Simon98,DasSarma96,Halperin96,Stormer96,Jain00,Halperin03,Smet98a,Murthy03}):

\begin{figure}
\includegraphics[width=3.5in]{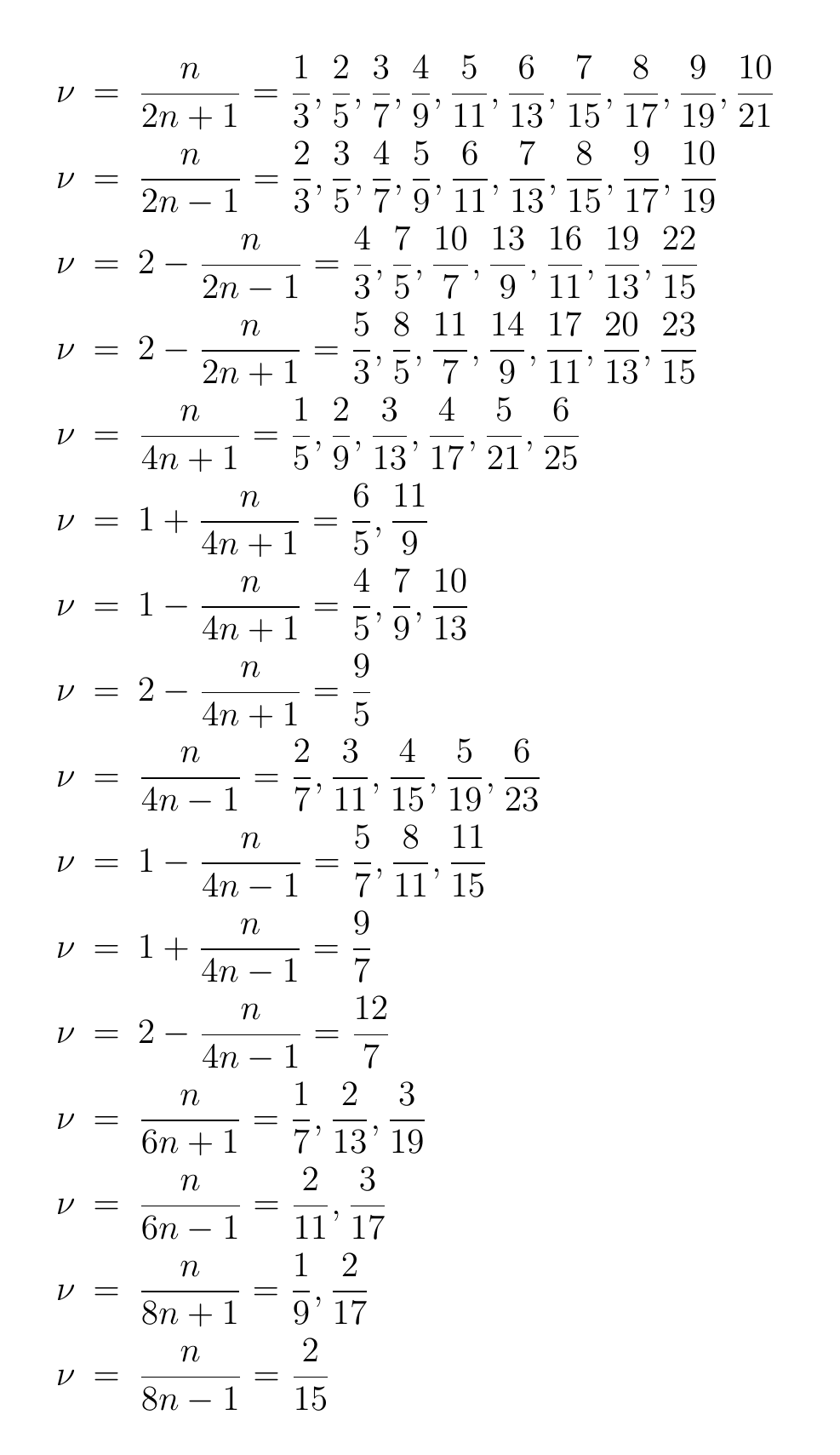}
\caption{The observed fractions in the lowest Landau level ($0\leq \nu \leq 2$) that correspond to IQHE of composite fermions\cite{Stormer-IQHE-FQHE,Pan02,Pan03}. Additionally, there exists evidence\cite{Stormer-IQHE-FQHE,Pan03} for $4/11$, $5/13$, $4/13$, $7/11$, $5/17$, $6/17$ (and indication also for $3/8$ and $3/10$\cite{Pan03}), which are not explicable as IQHE of composite fermions. Many of the above fractions have been seen only in the longitudinal resistance $R_{\rm xx}$; it is probable that the corresponding Hall plateaus will eventually be seen under sufficiently improved experimental conditions. The fractions for composite fermions with vorticity $2p=6$ and $2p=8$ appear only under somewhat elevated temperatures \cite{Pan02}, presumably above the melting temperature of the crystal ground state. The sequence $1-n/(2n+1)$ produces the same fractions as $n/(2n-1)$. It has been shown that the wave functions from the two methods are almost identical for fully spin polarized states, but the interpretation as $n/(2n-1)$ is crucial for an explanation of the spin physics of these states\cite{Wu93}; for example this is the only way to understand the spin singlet FQHE at 2/3, by analogy to the spin singlet state at $\nu^*=2$. Source: Stormer\cite{Stormer-IQHE-FQHE}.
\label{Fig-frac}}
\end{figure}

\begin{figure*}
\includegraphics[width=1.0\textwidth]{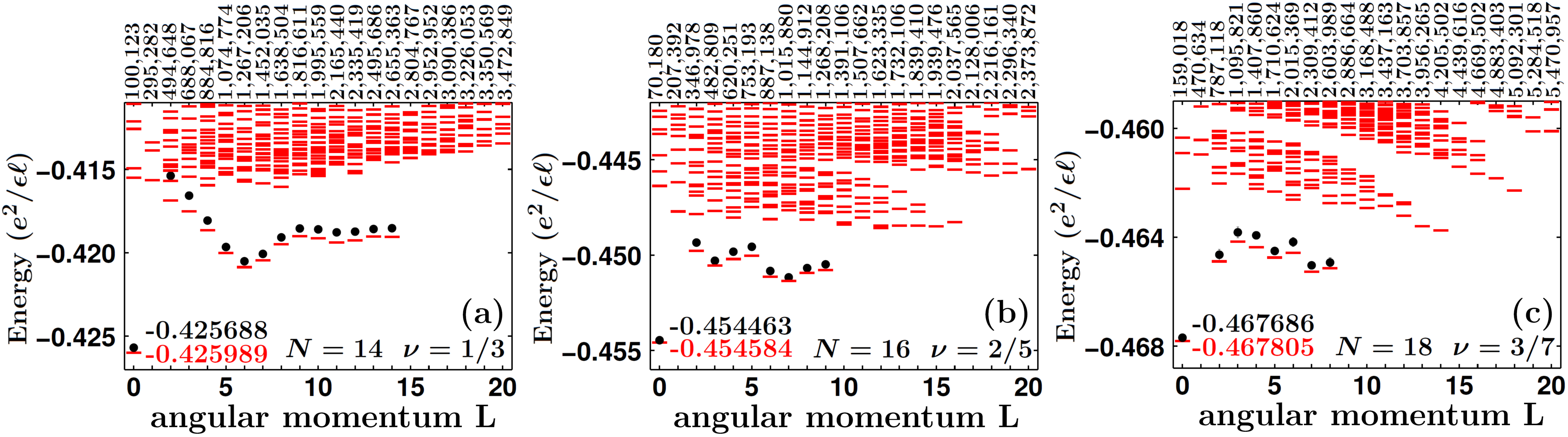}
\caption{This figure shows a comparison between the energies predicted by the CF theory (dots) and the exact Coulomb energies (dashes), both obtained without any adjustable parameters. The spectra shown are for (a) 14 electrons exposed to 39 flux quanta, (b) 16 electrons exposed to 36 flux quanta, and (c) 18 electrons exposed to $37$ flux quanta; these are finite size representations of the 1/3, 2/5 and 3/7 states. The dots show the Coulomb energies of the CF wave functions for the ground states and the neutral excitations (the CF wave function for the 1/3 ground state at $L=0$ coincides with Laughlin's wave function). The energies (per particle) are given in units of $e^2/\epsilon l$, and include the interaction with the uniform positively charged background. The spherical geometry\cite{Haldane83} is assumed, in which electrons move on the surface of a sphere in the presence of a radial magnetic field. The eigenstates are organized according to their total orbital angular momentum $L$, which is a good quantum number in this geometry. The Hilbert space is restricted to the lowest Landau level. Only the very low energy part of the spectrum is shown; the number of linearly independent multiplets in each $L$ sector is shown at the top. 
The CF energies can be further improved by incorporating CF Landau level mixing (compare, for example, the Fig.~6 of Ref.~\onlinecite{Balram13} with the above spectrum at 3/7). A large number of comparisons have been performed also at {\em arbitrary} filling factors in between the $n/(2n+1)$ states in the lowest Landau level (e.g. Ref.~\onlinecite{Balram13}), and show that the CF theory gives a faithful and accurate account of the lowest energy band seen in the exact Coulomb spectra\cite{missing}. This figure is taken from Ref.~\onlinecite{Balramunpub}; more comparisons can be found in Refs.~\onlinecite{Jain07}, \onlinecite{Jain98}, \onlinecite{Balram13}, and many references therein.}
\label{Figure1}
\end{figure*}

\begin{itemize}

\item The close correspondence between the FQHE plotted as a function of $B^*$ and the IQHE  
-- see Fig.~\ref{fqhe} (Ref.~\onlinecite{Stormer-IQHE-FQHE}).

\item Robust FQHE along the sequences $n/(2pn\pm 1)$\cite{Stormer-IQHE-FQHE,Pan02,Pan03} (63 of the 71 fractions observed in the lowest Landau level are explained as IQHE of composite fermions -- see Fig.~\ref{Fig-frac}).

\item Approximately linear $|B^*|$ dependence of the gaps at $n/(2pn\pm 1)$ (which is interpreted as the CF cyclotron energy, and produces an experimental measurement of the CF mass $m^*$) \cite{Du93,Manoharan94}.

\item Spin / valley polarizations of the FQHE states as a function of the Zeeman energy, and CF Landau level fan diagram \cite{Du95,Du97,Kukushkin99,Bishop07, Padmanabhan09,Feldman13}.

\item Dispersions of the neutral excitations (CF excitons) of $n/(2n+1)$ FQHE states\cite{Kukushkin09,Kang01,Rhone11}; sub-Zeeman energy spin reversed excitations \cite{Wurstbauer11}.

\item The CF Fermi sea at $\nu=1/2p$ \cite{Halperin93,Stormer95,PT93,Willett93,Kang93,Goldman94,Smet96,Smet98,Willett99,Smet99,Gokmen10,Kamburov12,Kamburov13}.

\item Spin polarization of the CF Fermi sea \cite{Kukushkin99,Melinte00}.

\item Shubnikov-de Haas oscillations of composite fermions \cite{Du94,Leadley94}.

\item Measurement of the CF cyclotron orbits by surface acoustic waves\cite{Willett93}, magnetic fousing \cite{Goldman94,Smet96}, and commensurability oscillations in periodic potentials \cite{Kang93,Smet98,Smet99,Willett99,Gokmen10,Kamburov12,Kamburov13}.

\item Cyclotron resonance mode of composite fermions \cite{Kukushkin02,Kukushkin07}.

\end{itemize}

The experiments confirm the central premise of the CF theory, namely the existence of weakly interacting fermions that experience an effective magnetic field $B^*$. The unusual stability of FQHE along the sequences $n/(2pn\pm 1)$ combined with the observation of the CF Fermi sea at $\nu=1/2p$ is an experimental proof that the FQHE at $n/(2pn\pm 1)$ is the IQHE of composite fermions. One may view the CF Fermi sea as emerging in the large $n$ limit of the $n/(2pn\pm 1)$ FQHE states\cite{Girvin}, just as the electronic state with many filled Landau levels merges smoothly into the electron Fermi sea.  Alternatively, one may understand the FQHE starting from the CF Fermi sea. As the filling factor is varied away from $\nu=1/2p$, first Shubnikov-de Haas oscillations in $B^*$ appear, which then develop into the CF-IQHE sequences $n/(2pn\pm 1)$; equivalently, composite fermions execute semiclassical cyclotron orbits at small $|B^*|$, which, with increasing $|B^*|$, become quantized into CF Landau levels and produce the CF-IQHE.  Experiments thus have shown that the $n/(2pn\pm 1)$ FQHE states and the $1/2p$ CF Fermi sea are not distinct phenomena but are inextricably linked by a common underlying physical origin.

The measurements of the cyclotron trajectories \cite{Willett93,Goldman94,Smet96,Kang93,Smet98,Smet99,Willett99,Gokmen10,Kamburov12,Kamburov13} with radius governed by $B^*$ have been considered direct observations of composite fermions.

The CF theory has also been corroborated by extensive quantitative calculations. Comparisons with exact results known for finite $N$ from numerical diagonalizations of the Coulomb Hamiltonian, such as those shown\cite{Balramunpub} in Fig.~\ref{Figure1} at 1/3, 2/5 and 3/7 (these represent the largest systems at these $\nu$ for which exact results are available), demonstrate that the CF theory predicts, with no adjustable parameters, the ground state energies to better than 0.1\% accuracy, and the dispersions of the neutral CF exciton to within a few percent accuracy. (These results establish the relation between the FQHE and the IQHE, a fundamental feature of the CF theory, at a microscopic level.) Such studies, combined with the Chern-Simons\cite{Lopez91,Halperin93} and Hamiltonian\cite{Murthy03} formulations of composite fermions, have enabled detailed predictions for the ground state energies \cite{Jain97}, gaps \cite{Scarola02}, neutral exciton dispersions\cite{Girvin85,Scarola00,Kukushkin09,Kang01,Rhone11,Murthy99,Mandal01}, spin polarization as a function of the Zeeman energy\cite{Park98,Davenport12}, spin wave dispersions\cite{Wurstbauer11}, surface acoustic wave absorption\cite{Halperin93,Stormer95,PT93,Willett93}, {\em etc}. These are in satisfactory agreement with the experimental measurements\cite{commex}. The quantitative studies of the FQHE via the CF theory are too numerous to recount here in a comprehensive fashion; for a survey of the literature in this context see Refs.~\onlinecite{Jain89,Lopez91,Halperin93,Stormer95,PT93,Jain07,Heinonen98,Jain98,Lopez98,Simon98,DasSarma96,Halperin96,Stormer96,Jain00,Halperin03,Smet98a,Murthy03}.

A comment is in order regarding the energy scales responsible for various phenomena\cite{commen}. The calculated gaps for the spin conserving excitations range from 0.1 $e^2/\epsilon l$ at 1/3 to $\sim$ 0.02 $e^2/\epsilon l$ at 7/15 (see Ref.~\onlinecite{Scarola02} and the references therein), and the Coulomb energy of the lowest spin reversed neutral excitation is $\sim -$0.005 $e^2/\epsilon l$ for the fully spin polarized 2/5 and 3/7 FQHE states\cite{Wurstbauer11}. The differences between the Coulomb energies (per particle) of differently spin polarized states, which determine the Zeeman energies at which transitions between them take place, range from $\sim 0.005$ $e^2/\epsilon l$ at 2/5 to $\sim 0.001$ $e^2/\epsilon l$ at 5/11 \cite{Park98}. The energy difference between the liquid and crystal states governing the physics of the re-entrant crystal observed\cite{Jiang90,Goldman90} between 1/5 and 2/9 is $\sim$0.0007 $e^2/\epsilon l$ per particle\cite{Archer13}. These theoretical numbers are subject to corrections when certain other aspects of the actual experimental systems are included in the calculation\cite{commex}, but they serve as a measure of the rather stringent accuracy required from a  quantitative theory of the FQHE.

What about the FQHE at $\nu\neq n/(2pn\pm 1)$, such as $4/11$, 5/13 and 3/8 \cite{Pan03}? These correspond to composite fermions at $\nu^*=4/3$, 5/3 and 3/2, and their explanation requires a treatment of the residual interaction between composite fermions in the second CF Landau level. This interaction has been determined from microscopic calculations within the CF theory (e.g. see Refs.~\onlinecite{Sitko96,Lee01}) and has a rather peculiar form, which has led to predictions of unconventional FQHE states of composite fermions at these fractions\cite{Wojs04,Mukherjee13,Mukherjee12}. As expected, the physics of these {\em fractional} QHE states of composite fermions, which occur at the second level of the CF hierarchy, is governed by much smaller energy scales than that of the nearby CF-IQHE states at 1/3 and 2/5, as indicated by rough estimations\cite{Wojs04,Mukherjee13,Mukherjee12} that show gaps on the order of $\sim$0.002 $e^2/\epsilon l$. 

The HH hierarchy theory does not explain the above listed experimental facts. In particular, according to the HH hierarchy mechanism, many observed fractions of the form $n/(2pn\pm 1)$  listed in Fig.~\ref{Fig-frac} would not be expected to occur in experiments, because they appear at very late generations of the HH hierarchy tree -- for example, 4/9 at the fourth, 5/11 at the 5th, and ... 10/21 at the 10th generation (see the discussion in Sec.~\ref{stability}). Quantitative calculations for various experimentally relevant quantities, such as ground state energies, gaps, neutral exciton dispersions, spin phase diagrams, etc., have not yet been carried out within the HH hierarchy theory. Comparisons with exact diagonalization results have been performed only for very small systems and have been inconclusive\cite{Greiter2}.

\section{Universal aspects} 
\label{secIII}

The two theories produce the same values for the local charge and braid statistics ($e^*$ and $\theta^*$) of the quasiparticle excitations\cite{commfs,Su86}.  They belong to equivalent K-matrix classifications of the effective Chern-Simons effective action\cite{CS}, which encodes information on $e^*$ and $\theta^*$ and has been used as a starting point for an effective description of certain low-energy universal properties of a given FQHE state\cite{universal}. These facts, by themselves, do not imply equivalence of two theories, for reasons explained in the Introduction. 

Notably, the central role in the CF theory is played by a {\em new} topological quantum number, namely the CF vorticity \cite{topo} $2p$, which is measurable, has been measured, and is responsible for much of the observed phenomenology.  (While $e^*$ and $\theta^*$ do appear in the CF theory, their consideration is not required for the explanation of the FQHE, the CF Fermi sea, or any of the other phenomena listed in Section \ref{secII}.) The CF vorticity is intrinsic to the definition of composite fermions -- it is what makes them topologically distinct from electrons. The vorticity of composite fermions manifests directly through the magnetic field $B^*=B-2p\rho\phi_0$ experienced by them; as composite fermions move about, the Berry phase due to the bound vortices partly cancel the Aharonov-Bohm phase due to the external magnetic field to produce the effective field $B^*$\cite{Jain07}. Appendix \ref{secIV} shows that the vorticity of composite fermions is more general than the $e^*$ and $\theta^*$ of the excitations.

\section{Concluding remarks}
\label{conclusion}

We have noted that the CF and the HH hierarchy theories postulate distinct mechanisms for the origin of the FQHE and make substantially different predictions for experiments\cite{conflate}. Experiments have verified numerous nontrivial consequences of the CF theory, such as the existence of composite fermions, the effective magnetic field ($B^*$), the CF Fermi sea, the prominent fractional sequences $n/(2pn\pm 1)$, the spin physics, various kinds of excitations, and a host of other phenomena. The CF theory unifies the incompressible and compressible states, and also the FQHE and the IQHE. 

We end with a mention of some open problems. Further work would be needed to settle the nature of the states such as 4/11, 5/13, and 3/8, which have been predicted to have an unconventional origin\cite{Wojs04,Mukherjee12,Mukherjee13}; there is hardly any doubt, however, that they are some kinds of FQHE states of interacting composite fermions. Another feature in need of convincing experimental confirmation is the nature of the lowest Landau level crystal phase at low fillings, which has been predicted to contain a series of CF crystals\cite{Yi98,Narevich01,Chang05,Archer13}. As for the second Landau level, exact diagonalization studies have shown that the FQHE at filling factors $2+n/(4n\pm 1)$ is well described as IQHE of composite fermions carrying four vortices\cite{Wojs09}, but the physical origin of the states at $2+n/(2n\pm 1)$, including\cite{Xia04,Kumar10} $2+3/8$ and $2+6/13$, is currently being debated. The observation of reentrant IQHE\cite{Eisenstein02,Xia04,Kumar10} in the second Landau level indicates nearby competing phases, which complicates the analysis. The FQHE at 5/2\cite{Willett87} has been modeled in terms of a chiral p-wave paired state of composite fermions \cite{Moore91,Read00}. This state has been predicted to support Majorana modes\cite{Read00}, i.e. half-composite fermions trapped in the Abrikosov vortices, which are believed to obey non-Abelian braid statistics and are a subject of current experimental study\cite{Willett10}. Another unresolved issue is the nature of the FQHE edge. Experimental measurements of the edge exponents\cite{Chang03} have, so far, not produced the precisely quantized values predicted by the effective K-matrix approach, for reasons not yet fully understood. There is currently much excitement regarding the observation of composite fermions and FQHE in materials with multiple valleys, such as AlAs quantum wells\cite{Bishop07,Padmanabhan09}, graphene\cite{Feldman13,Xu09,Bolotin09,Dean11,Feldman12} and Si surfaces\cite{Kott14};  combined with spin, these may allow a study of the SU(4) limit. One can also look forward to a realization of the FQHE physics in bosonic systems\cite{Lin09,Gemelke10} as well as at the surface of topological insulators. Investigations of these and other issues hold the promise of important future discoveries.

{\bf Acknowledgments:} I am grateful to H. L. Stormer for permission to use Fig.~\ref{fqhe} and for the list of observed fractions given in Fig.~\ref{Fig-frac}. I am indebted to B. I. Halperin for detailed and insightful comments on an earlier version of this manuscript, which have been incorporated into the article, and for his kind encouragement.  I also thank J. R. Banavar, S. Das Sarma, G. Murthy, D. S. Weiss, A. W\'ojs, and Y.-H. Wu for useful discussions in this context.  Financial support from DOE under Grant No. DE-SC0005042 is gratefully acknowledged.

\begin{appendix}

\section{Composite fermions $\neq$ Laughlin quasiparticles}
\label{ap:LQP}

The HH hierarchy and the CF theories have one common point, namely the Laughlin ground state at $\nu=1/m$. This Appendix shows that the two begin to diverge as soon as the filling factor is increased, and that the divergence reflects a structural difference between the two theories. In particular, creating Laughlin quasiparticles (LQPs) results in states with nonzero amplitude in up to very high CF Landau levels.

{ The trial wave function for a} single LQP at the origin for $\nu=1/m$ ($m=2p+1)$ is given by
\begin{equation}
\Psi^{\rm LQP} = 
e^{ -\sum_j |z_j|^2/4} 
\left(\prod_l  2 \frac{\partial}{\partial z_l} \right) 
\prod_{j<k} (z_j - z_k)^m, 
\label{LQP1}
\end{equation}
The trial wave function for a single composite fermion in the second CF Landau level at the origin, labeled ``CF quasiparticle" (CF-QP), is given by\cite{Jain89b} 
\begin{eqnarray}
&&\Psi^{\rm CF-QP} 
= \exp \left[ -\frac14 \sum_j |z_j|^2 \right]  \times \nonumber\\
&& {\cal P}_{\rm LLL} 
\left|
\begin{array}{ccc}
\bar{z}_1 & \bar{z}_2 & \ldots\\
1 & 1 & \ldots\\
z_1 & z_2 & \ldots\\
\vdots & \vdots&\ldots\\
z_1^{N-2} & z_2^{N-2} & \ldots
\end{array}
\right| \prod_{j<k} (z_j - z_k)^{2p} 
\label{JQP1}
\end{eqnarray}
where the determinant is the wave function with one particle in the second Landau level, and the lowest Landau level projection is accomplished by the replacement $\bar{z}_j\rightarrow 2\partial/\partial z_j$.  The wave functions $\Psi^{\rm LQP}$ and $\Psi^{\rm CF-QP}$ are not identical;  explicit calculation for Coulomb interaction in the lowest Landau level has shown \cite{Jeon03b} that the CF quasiparticle in Eq.~\ref{JQP1} has $\sim$15\% lower energy than the LQP of Eq.~\ref{LQP1}. 

Is this quantitative difference an indication of a qualitatively different underlying structure? To gain an insight into this question let us analyze the LQP from the perspective of the CF theory. In the CF theory, the wave function of a composite fermion in the $(n+1)$th CF Landau level contains derivatives with powers up to $[\partial/\partial z_j]^n$. The CFQP in Eq.~\ref{JQP1} has precisely one composite fermion in the second CF Landau level, and none in higher ones. The LQP in Eq.~\ref{LQP1} has no composite fermions in the third and higher CF Landau levels, but it has a non-zero probability of containing {\em many} composite fermions in the second CF Landau level. That the difference between the LQPs and composite fermions is qualitative and structural becomes indisputable as more LQPs are created. The state with two LQPs at ${\eta}_1$ and ${\eta}_2$
\begin{equation}
e^{ -{1\over 4}\sum_j |z_j|^2} 
\prod_l \left( 2 \frac{\partial}{\partial z_l} -\bar{\eta}_1\right) \left( 2 \frac{\partial}{\partial z_l} -\bar{\eta}_2\right) 
\prod_{j<k} (z_j - z_k)^m, 
\label{LQP2}
\end{equation}
has composite fermions occupying the {\em third} CF Landau level as well. (For nearby LQPs, the energy of this state is substantially higher than that of the state with two nearby composite fermions occupying the second CF Landau level \cite{Jeon03b}.)  Analogously, for $N_{\rm LQP}$ LQPs, the wave function $\Psi^{\rm LQP}_{1/m}( \vec{r}_j ;  \vec{\eta}_{\mu} )$ in Eq.~\ref{LQP} has a nonzero occupation of the lowest $N_{\rm LQP}+1$ CF Landau levels. Adding LQPs is thus very different from filling the second CF Landau level.  To reach the 2/5 daughter, $N_{\rm LQP}=N/2$ LQPs must be created, which produces a state that has amplitude extending up to $\sim N/2$ excited CF Landau levels. This is to be contrasted with the CF description of the 2/5 state as lowest two filled CF Landau levels.

\section{Composite-fermion vorticity}
\label{secIV}

A distinctive feature of the CF theory is to identify a new topological quantum number, namely the CF vorticity, which is intrinsic to the definition of the composite fermions themselves and manifests through the effective field $B^*=B-2p\rho\phi_0$. The following facts demonstrate that the composite fermions and their vorticity are { more general} than the local charge and braid statistics ($e^*$ and $\theta^*$) of the excitations.

(i) We first note that we can derive $e^*$ and $\theta^*$ from composite fermions \cite{Jain07}. One may ask: ``The existence of objects with fractional $e^*$ and $\theta^*$ can be deduced from general principles \cite{commfs}, but what are these objects?" The CF theory tells us what they really are: they are isolated composite fermions in an otherwise empty CF Landau level or missing composite fermions in an otherwise filled CF Landau level. These are sometimes referred to as ``CF quasiparticles" and ``CF quasiholes" when viewed relative to the uniform $\nu^*=n$ ``vacuum." This description has been demonstrated to give a precise account of the excitations of all $n/(2pn\pm 1)$ FQHE states -- see, e.g., Refs.~\onlinecite{Jain07,Balram13}.  (Laughlin's ``quasihole" of the $1/m$ state is identical to the CF quasihole for this state. The ``quasielectrons" for $\nu=1/m$ in Refs.~\onlinecite{Haldane1,Haldane2,Haldane3}, or for the other CF states in Ref.~\onlinecite{Hansson09,Surosa1}, also precisely match those of the CF theory.)
The local charge (i.e. charge excess relative to the uniform ``vacuum" FQHE state) of a CF quasiparticle at $\nu=n/(2pn\pm 1)$ is the sum of the charges of an electron and $2p$ vortices; recognizing that the charge of a vortex is simply $\nu$, the local charge is given by $e^*=-1+2p\nu=-1/(2pn\pm 1)$.  A Berry phase calculation
\cite{Leinaas99,Jeon03} shows that the braid statistics of the CF quasiparticles is well defined provided their spatial overlap is negligible, and is given by the product of the vorticity and the local charge, $\theta^*=2p\times {1\over 2pn\pm 1}$ (see Ref.~\onlinecite{Jain07} for further details). Both $e^*$ and $\theta^*$ thus inherit their quantized values from the quantized CF vorticity $2p$.

The existence of composite fermions and their vorticity, on the other hand, does not follow from the knowledge of $e^*$ and $\theta^*$. In particular, the general arguments outlined in Footnote \onlinecite{commfs} that give us $e^*$ and $\theta^*$ for a given $\nu$ do not give any indication of the existence of composite fermions. The CF theory thus contains whatever follows from $e^*$ and $\theta^*$, but also much that does not.

(ii) Viewed solely through their $e^*$ and $\theta^*$ quantum numbers, as would be the case if we did not know about composite fermions, it would seem that the excitations of different FQHE states are fundamentally distinct, producing $\sim$70 distinct particles. The CF theory reveals that they are all the same. Furthermore, they are also identical to the particles forming the ground states. The same composite fermions are used to build the ground states, the charged excitations, the neutral excitations, and multiple excitations for {\em all} states of the form $n/(2pn\pm 1)$ with a given $2p$. Instead of $\sim$70 fractionally charged anyons\cite{Leinaas77}, it is thus sufficient to work with only a few flavors of composite fermions with different vorticity. Different values of $e^*$ and $\theta^*$ occur simply because the charge of a vortex ($\nu$) depends on the filling factor of the background FQHE state.

(iii) Composite fermions and their vorticity are well defined over a broader region than their fractional $e^*$ and $\theta^*$. We illustrate with some examples:

\begin{itemize}

\item Composite fermions in a filled CF Landau level do not have any $e^*$ or $\theta^*$, as they are part of the ``vacuum". Their vorticity is well defined, however -- the resulting $B^*$ is what gave us the filled CF Landau level state in the first place. 

\item Imagine only a few composite fermions in a CF Landau level, i.e. a few CF quasiparticles.  They have well defined local charge $e^*$ and braid angle $\theta^*$, but only provided they have negligible spatial overlap with one another. Explicit Berry phase calculations \cite{Leinaas99,Jeon03} for the CF quasiparticles of the 1/3 and 2/5 FQHE states show that they must be farther than $\sim$10 magnetic lengths in order for $\theta^*$ to have a well defined value. For other FQHE states the CF quasiparticles have even larger sizes, requiring larger separations to ensure a well defined $\theta^*$. Furthermore, the Landau level mixing, always present, introduces corrections to $\theta^*$ that decay only as a power law in the distance between the quasiparticles \cite{Simon08}. Detailed calculations have also demonstrated that the interaction between the CF quasiparticles is weak and often {\em attractive} at short distances \cite{Lee01}, implying that there exists no energy barrier keeping them far apart from one another. 

In contrast, the description in terms of CF quasiparticles remains well defined and accurate even when they are nearby and overlapping. This is demonstrated by the accuracy of the CF theory in describing even small systems containing multiple CF quasiparticles and / or CF quasiholes \cite{Jain07,Balram13}.

\item As we start populating a CF Landau level with more composite fermions, at some point, it is not possible, even in principle, to keep all CF quasiparticles away from one another, and $e^*$ and $\theta^*$ cease to be meaningful quantum numbers. However, composite fermions and their $B^*$ remain sharply defined all the way to the filled CF Landau level state, and beyond. It is thus the vorticity (or $B^*$) and the exchange statistics of composite fermions that are responsible for incompressibility and FQHE at $n/(2pn\pm 1)$.

\item Last, the vorticity of composite fermions manifests itself, through an effective $B^*$, also in compressible regions (e.g. the 1/2 CF Fermi sea), which cannot support, even as a matter of principle, excitations with well defined local charge and braid statistics. 

\end{itemize}

(iv) One can ask what relevance these quantum numbers have to experiments.  
Many of the experimental facts discussed in Sec.~\ref{secII} are direct consequences of $B^*$ and hence of the CF vorticity. The vorticity of composite fermions has been determined directly also in experiments that measure the cyclotron orbits of the objects responsible for transport \cite{Willett93,Kang93,Goldman94,Smet96,Smet98,Smet99,Willett99,Gokmen10,Kamburov12,Kamburov13} which are seen to correspond to the effective field $B^*$. Shot noise and interference experiments have been designed for detecting the $e^*$ and $\theta^*$ quantum numbers of the excitations (e.g., Refs.~\onlinecite{dePicciotto97,Willett10}).

\end{appendix}


\begin{thebibliography}{99}
\bibitem{universal} For example, it predicts universal behavior for the FQHE edge, which has been a subject of much experimental investigation\cite{Chang03}.
\bibitem{comm-1m} The fractions $1/(2p+1)$, which lie at the boundary separating regions spanned by composite fermions with different vorticity, have a dual interpretation: either as filling factor $1$ of composite fermions carrying $2p$ vortices, or as filling factor ``$-1$" of composite fermions carrying $2(p+1)$ vortices. Both produce similar wave functions, but the values of the gaps suggest the former interpretation as being more natural. That is the reason why the sequences $n/(2pn -1)$ begin with $n=2$ in Fig.~\ref{cfsequences}.
\bibitem{commex} 
The agreement with experiments is less accurate than that with exact diagonalization results because the experimental numbers are also affected by finite width corrections, Landau level mixing and disorder, for which our understanding is not as precise.
\bibitem{commen} As a reference, the energy of two electrons in the lowest Landau level in the state with unit relative angular momentum is $\sim$0.44 $e^2/\epsilon l$.
\bibitem{Greiter2} Ref.~\onlinecite{Greiter} has compared the HH hierarchy wave functions with the 2/5 Coulomb ground state for 6 and 8 particles, and with the 3/7 Coulomb ground state for 6 particles. These systems are too small for a convincing test of the wave functions, as they have only 3, 8 and 2 uniform $L=0$ eigenstates, respectively.  (In contrast, the 16 particle 2/5 system and the 18 particle 3/7 system shown in Fig.~\ref{Figure1} have, respectively, 70180 and 159018 eigenstates in the $L=0$ sector.)  Furthermore, the actual Coulomb state of 6 particles at total flux 9, which was assigned the filling factor 3/7 in this work, is more appropriately viewed as the 6 particle 2/3 state, and is accurately represented by the hole counterpart of the 4 particle 1/3 state. This illustrates how studies of small systems can fail to discriminate between wave functions based on very different physics, necessitating studies of larger systems and comparisons with experiments for a more substantive test of the theory.
\bibitem{commfs} The fact that the CF and the HH hierarchy theories produce the same values for $e^*$ and $\theta^*$, despite their different physics for the origin of the FQHE, is less surprising than it might at first seem. The mere assumption of incompressibility at a given fractional filling $\nu$ imposes stringent constraints on $e^*$ and $\theta^*$ without regard to the mechanism responsible for incompressibility, and both the CF and the HH hierarchy theories obtain the simplest allowed values for them. The logic is as follows. Laughlin's method of adiabatic flux insertion at some location in an incompressible state at filling factor $\nu$ produces a localized object with fractional charge excess of $\nu$ (in units of the electron charge), and one can give strong arguments that the bound state of a fractional charge $\nu$ and a unit flux quantum obeys fractional braid statistics\cite{Halperin84}. (Note that this derivation of fractional charge by adiabatic flux insertion does not depend on the origin of incompressibility.) In general the charge $\nu$ object is not an elementary excitation, however. Requiring that an integer number of elementary excitations can be combined to produce a charge $\nu$ excitation and another integer number of them produce an electron imposes further constraints on the allowed values of $e^*$ and $\theta^*$ of the elementary excitations. For the fractions $\nu=n/(2pn\pm 1)$ this shows that the allowed values of local charge are $e^*=1/[r(2pn\pm 1)]$, where $r$ is an integer; integer multiples of this can produce both $\nu$ and 1. This closely follows Ref.~\onlinecite{Su86}, which also contains a discussion of general fractions. The CF and HH hierarchy theories obtain the simplest allowed values: $e^*=1/(2pn\pm 1)$ and the corresponding $\theta^*=2p/(2pn\pm 1)$.
\bibitem{topo} The vortices are topological because a closed loop around a vortex produces a phase of $2\pi$ independent of the size or the shape of the loop.
\end{thebibliography}
\end{document}